\documentclass[floats,floatfix,showpacs,amssymb,prd,nofootinbib,superscriptaddress,aps,notitlepage,twocolumn]{revtex4-1}

\usepackage{graphicx}
\usepackage[utf8]{inputenc}
\usepackage{graphicx, epsfig, amssymb} 
\usepackage{amsmath, amsfonts}
\usepackage{bm} 

\usepackage[linktocpage]{hyperref}
\usepackage[caption=false]{subfig}
\usepackage[usenames]{color}

\def\be{\begin{equation}}
\def\ee{\end{equation}}

\newcommand{\bea}{\begin{eqnarray}}
\newcommand{\eea}{\end{eqnarray}}
\newcommand{\ben}{\begin{enumerate}}
\newcommand{\een}{\end{enumerate}}
\newcommand{\bi}{\begin{itemize}}
\newcommand{\ei}{\end{itemize}}

\newcommand{\nn}{\nonumber}

\def\ga{\mathrel{\raise.3ex\hbox{$>$\kern-.75em\lower1ex\hbox{$\sim$}}}}
	\def\la{\mathrel{\raise.3ex\hbox{$<$\kern-.75em\lower1ex\hbox{$\sim$}}}}

\def\l{\left}
\def\r{\right}
\def\be{\begin{equation}}
\def\ee{\end{equation}}

\def\I_M{{I_{\scriptscriptstyle M\times M}}}

\def\be{\begin{equation}}
\def\ee{\end{equation}}
\def\bea{\begin{eqnarray}}
\def\eea{\end{eqnarray}}
\newcommand{\beq}{\begin{eqnarray}}
\newcommand{\eeq}{\end{eqnarray}}
\def\pa{\partial}

\def\lam{\lambda_{lm}}

\newcommand{\olm}[1]{#1_{\omega l m}}

\newcommand{\beqal}{\begin{eqnarray}\label}
\newcommand{\none}{\end{eqnarray}}
\newcommand{\beqa}{\begin{eqnarray}}
\newcommand{\eeqa}{\end{eqnarray}}

\newcommand{\siggeo}{\sigma_{\text{geo}}}

\begin{document}\title{\large Absorption of planar massless scalar waves by Kerr black holes}

\author{Caio F. B. Macedo}\email{caiomacedo@ufpa.br}
\affiliation{Faculdade de F\'{\i}sica, Universidade 
Federal do Par\'a, 66075-110, Bel\'em, Par\'a, Brazil.}

\author{Luiz C. S. Leite}\email{luiz.leite@icen.ufpa.br}
\affiliation{Faculdade de F\'{\i}sica, Universidade 
Federal do Par\'a, 66075-110, Bel\'em, Par\'a, Brazil.}

\author{Ednilton S. Oliveira}\email{ednilton@ufpa.br}
\affiliation{Faculdade de F\'{\i}sica, Universidade 
Federal do Par\'a, 66075-110, Bel\'em, Par\'a, Brazil.}

\author{Sam R. Dolan}\email{s.dolan@sheffield.ac.uk}
\affiliation{Consortium for Fundamental Physics,
School of Mathematics and Statistics,
University of Sheffield, Hicks Building, Hounsfield Road, Sheffield S3 7RH, United Kingdom.}

\author{Lu\'is C. B. Crispino}\email{crispino@ufpa.br}
\affiliation{Faculdade de F\'{\i}sica, Universidade 
Federal do Par\'a, 66075-110, Bel\'em, Par\'a, Brazil.}

\begin{abstract}	
We consider planar massless scalar waves impinging upon a Kerr black hole, for general angles of incidence. We compute the absorption cross section via the partial wave approach, and present a gallery of results. In the low-frequency regime, we show that the cross section approaches the horizon area; in the high-frequency regime, we show that the cross section approaches the geodesic capture cross section.  
In the aligned case, we extend the complex angular momentum method to obtain a `sinc' approximation, which relates the regular high-frequency oscillations in the cross section to the properties of the polar null orbit. In the non-aligned case, we show, via a semi-analytic approximation, that the reduction in symmetry generates a richer, less regular absorption cross section. We separate the absorption cross section into corotating and counterrotating contributions, showing that the absorption is larger for counterrotating waves, as expected.
\end{abstract}


\date{\today}

\maketitle


\section{Introduction}
Black holes (BHs) \cite{chandra} are among the most intriguing predictions of general relativity (GR) \cite{wald}. 
In electrovacuum, BHs are described by the Kerr-Newman family of solutions, which are governed by just three parameters: mass, spin and electric charge. The no-hair theorem \cite{lrr-2012-7} implies that black holes cannot support additional degrees of freedom, which suggests that in essence black holes are rather simple objects, even if their astrophysical environments are likely to be extremely complex. 

BHs are believed to populate the galaxies \cite{Narayan:2005ie}. Accumulated evidence that supermassive rotating BHs reside at the center of active galactic nuclei has had a profound impact \cite{Ferrarese:2004qr}. Some of these rotating BHs are expected to be spinning very close to their upper rotating limit \cite{berti,Risaliti:2013cga, Walton:2013}, and so the phenomenology around rotating BHs is of major importance to astrophysics. This motivates careful study of the nature and observational consequences of the Kerr metric, which describes a rotating BH in GR.  An improved understanding of the Kerr BH will also help us to understand more complicated structures, such as rotating BHs in modified theories of gravity \cite{Clifton:2011jh, Kleihaus:2011tg,Pani:2011gy,Yagi:2012ya}.

One aspect of Kerr phenomenology is the absorption and scattering of particles by BHs. Particles are described by (quantum and fluid) field theories, and so the absorption and scattering of particles is naturally related to the absorption and scattering of fields, which may have spin, mass and charge. Spinless (i.e.~scalar) fields are particularly important, both as a particle model (e.g., for pseudoscalar mesons) and as a first model for bosonic fields with spin (e.g., the electromagnetic field). In theories that seek to go beyond the Standard Model, light scalar bosons may play important roles; for instance, in axiverse models \cite{Arvanitaki:2009fg,Arvanitaki:2010sy} and in scalar field dark matter models \cite{Boehm:2003hm}. The discovery of a Higgs-like particle by the ATLAS and CMS collaborations has given extra motivation to the study of scalar fields \cite{Higgs:2012gk}.  

The absorption and scattering cross sections of planar waves by black holes have been extensively studied. Recently, a unified picture of the scattering of massless planar waves by Schwarzschild BHs was presented \cite{Crispino:2009xt}. This builds upon the work of many authors in investigating the absorption cross section of planar waves by the Schwarzschild black hole \cite{Fabbri:1975sa,Sanchez:1977si,Gubser:1997cm,Doran:2005vm,Crispino:2007qw}. Various studies have also been made considering charged black holes as central scatterers \cite{Jung:2005mr,Crispino:2009ki,Crispino:2008zz,Oliveira:2011zz}. Despite their physical relevance, rotating BHs have received less attention in the literature, with several notable exceptions \cite{Glampedakis:2001cx,dolanthesis,Dolan:2008kf}. Absorption and scattering cross sections by acoustic BH analogues have also been recently investigated \cite{Crispino:2007zz,Dolan:2009zza,Dolan:2011ti,Oliveira:2010zzb}.

In this work, we analyze the absorption cross section of a planar massless scalar wave impinging upon a Kerr BH, giving emphasis to the general case in which the direction of incidence is not aligned with the axis of rotation. The paper is ordered as follows. In Sec.~\ref{sec:scalarfield} we describe the separation of variables for the massless scalar field in the Kerr spacetime in Boyer-Lindquist coordinates, and the physical boundary conditions for planar wave scattering. In Sec.~\ref{sec:abs} we give expressions for the absorption cross section in the Kerr spacetime, and identify the co and counterrotating contributions. We describe the low{-} and high-frequency regimes, and present an asymptotic formula for the absorption cross section arising from the complex angular momentum method. In Sec.~\ref{sec:results} we present a selection of numerical results, considering different values of the incident angle, and of the BH rotation parameter. We conclude with a discussion in Sec.~\ref{sec:conclusion}. Throughout, we use natural units ($c=G=\hbar=1$), and the metric signature ($+,-,-,-$).

\section{Scalar field in the Kerr spacetime}
\label{sec:scalarfield}
In the standard Boyer-Lindquist coordinate system  ($t,r,\theta,\varphi$), the Kerr BH is described by the line element \cite{kerr}
\bea
ds^2&&=\l(1-\frac{2Mr}{\rho^2}\r)dt^2+\frac{4Mar\sin^2\theta}{\rho^2}dt\,d\varphi -\frac{\rho^2}{\Delta}dr^2 \nn\\
&&-\rho^2d\theta^2-\l(r^2+a^2+\frac{2Mra^2\sin^2\theta}{\rho^2}\r)\sin^2\theta d\varphi^2,
\eea
in which $\Delta =r^2-2Mr +a^2$ and $\rho^2 = r^2+a^2\cos^2\theta$ {\cite{caio_erratum}}. From the asymptotic behavior, one may infer that $M > 0$ is the mass of the Kerr BH and $a \ge 0$ its angular momentum per unit mass ($a=J/M$). Here we restrict ourselves to the regime in which the Kerr metric describes a BH spacetime, i.e., $a\leq M$. For $a<M$ the Kerr BH has two distinct horizons. The inner (Cauchy) horizon is at $r_{-}~=~M-\sqrt{M^2-a^2}$ and the outer (event) horizon is at $r_{+}~=~M+\sqrt{M^2-a^2}$. If $a~=~M $ we have an extreme Kerr BH with an event horizon at $r_{+}~=~r_{-}~=~M$. The case $a > M$ corresponds to a naked singularity.

A massless scalar field $\Phi(x^\mu)$ in a curved background is governed by
\be
\frac{1}{\sqrt{-g}}\pa_\mu(\sqrt{-g}g^{\mu\nu}\pa_\nu \Phi)=0,
\label{scalar}
\ee
where $g_{\mu\nu}$ are the covariant metric components of the Kerr spacetime, $g$ the metric determinant and $g^{\mu\nu}$ are the contravariant metric components. Here we shall be interested in monochromatic wave-like solutions of Eq.~\eqref{scalar}, which can be obtained by separation of variables \cite{carter,brill}, so that we may write:
\be
\Phi= \frac{\olm{U}(r)}{\sqrt{r^2+a^2}} \olm{S}(\theta) e^{i m \varphi -i \omega t}.
\ee
The functions $\olm{S}(\theta)$ are the standard oblate spheroidal harmonics \cite{spheroid}, which will be normalized according to
\be
2 \pi \int d\theta \sin\theta ~ |\olm{S}(\theta)|^ 2=1.
\ee
The radial functions $\olm{U}(r)$ obey the following differential equation
\be
\l(-\frac{d^2}{dx^2}+\olm{V}(x)\r)\olm{U}(x)=\omega^2\olm{U}(x),
\label{radial}
\ee
with an effective potential given by
\bea	
\olm{V}(x)=-\frac{1}{\l(r^ 2+a^ 2\r)^2}[m^2a^2-4Mma\omega r&&
+ \nn\\
-\Delta(\lam+\omega^2a^2)]+\Delta\frac{\Delta+2r(r-M)}{\l(r^2+a^2\r)^3}
-\frac{3r^2\Delta^2}{\l(r^2+a^2\r)^4}.&&
\label{efpot}
\eea
In Eq.~(\ref{radial}) we made use of the tortoise coordinate $x$ of the Kerr spacetime, defined through 
\be
dx=\frac{r^2+a^2}{\Delta}dr. 
\ee
The constants $\lam$ are the eigenvalues of the oblate spheroidal harmonics (cf., e.g., Ref.~\cite{Macedo:2012zz}). The independent solutions of Eq.~\eqref{radial} are usually labeled as $in$, $up$, $out$ and $down$ (see, e.g., Ref.~\cite{Ottewill:2000qh}). Here we will be interested in the $in$ modes, since they characterize purely incoming waves from the past null infinity, obeying the following boundary conditions
\be
\olm{U}(x)\sim\l\{
\begin{array}{c l}
	\olm{\mathcal{A}}R_I+\olm{\mathcal{R}} {R_I}^*& (x/M\rightarrow \infty),\\
	\olm{\mathcal{T}} R_{II} & (x/M\rightarrow -\infty).
\end{array}\r.
\label{inmodes}
\ee
The functions $R_I$ and $R_{II}$ are given by
\bea
R_I&=&e^{-i \omega x} {\sum^N_{j=0}} \frac{A^j_\infty}{r^j},
\label{eq:expansion1}\\
R_{II}&=&e^{-i \tilde{\omega} x} \sum_{j=0}^N (r-r_+)^jA^j_{r_+},
\label{eq:expansion2}
\eea
where $\tilde{\omega}~\equiv~\omega - m a/(2M r_+)$, and the coefficients $A^j_{\infty}$ and $A^j_{r_+}$ are obtained by requiring that the functions $R_I$ and $R_{II}$ are solutions of the differential equation \eqref{radial} far from the BH and close to the outer horizon, respectively. The coefficients $\olm{\mathcal{R}}$ and $\olm{\mathcal{T}}$ are related to the reflection and transmission coefficients, respectively, and obey the following relation
\be
\l|\frac{\olm{\mathcal{R}}}{\olm{\mathcal{A}}}\r|^2=1-\frac{\tilde{\omega}}{\omega}\l|\frac{\olm{\mathcal{T}}}{\olm{\mathcal{A}}}\r|^2.
\ee
When $\omega \tilde{\omega}<0$, it follows that $|\olm{\mathcal{R}}|^2>|\olm{\mathcal{A}}|^2$, due to a phenomenon known as superradiance \cite{Press:1972zz}.

\section{Absorption cross section}
\label{sec:abs}
The partial absorption cross section of an asymptotic plane scalar wave propagating in the direction $\mathbf{n}=(\sin \gamma,0, \cos\gamma)$ is given by \cite{dolanthesis,Dolan:2008kf}
\be
\sigma_{lm}=\frac{4 \pi^2}{\omega^2}|\olm{S}(\gamma)|^2
\Gamma_{\omega l m} ,
\label{absorption}
\ee
where the transmission factors are
\be
\Gamma_{\omega l m} = 
\l(1-\l|\frac{\olm{\mathcal{R}}^{in}}{\olm{\mathcal{A}}^{in}}\r|^2\r),
\label{eq:trans}
\ee
and the total absorption cross section is
\be
\sigma=\sum_{l=0}^\infty\sum_{m=-l}^l\sigma_{lm}.
\label{eq:total-abs}
\ee
The cross section is invariant under $\gamma \rightarrow \pi - \gamma$. 
When the direction of incidence is parallel to the spin axis of the BH ($\gamma = 0$), we have $\sigma_{lm} = 0$ for $m \neq 0$. When the direction of incidence lies in the equatorial plane of the BH ($\gamma~=~90$ degrees) we have $\sigma_{lm}~=~0$ for odd values of $l+m$, because $\olm{S}(\pi/2)~=~0$ in this case. 
The total absorption cross section can be decomposed in the following way:
\be
\sigma=\sigma^+ +\sigma^-,
\ee
where 
\be
\sigma^{\pm}=\frac{1}{2}\sum_{l=0}^\infty\sigma_{l0}+\sum_{l=1}^\infty\sum_{m=1}^{l}\sigma_{l, \pm m}.
\label{eq:co-counter}
\ee
In this way we may separate the absorption cross  section into corotating ($\sigma^{+}$) and counterrotating ($\sigma^{-}$) contributions. 

\subsection{Low-frequency regime}
In the low-frequency regime, it has been shown that the absorption cross section for stationary BHs equals the area of the BH event horizon \cite{Higuchi:2001si,Higuchi:2001sib}. This result is quite general and does not depend on the direction of the incident wave. We have checked our numerical results in this limit, computing numerically the absorption cross section for low values of $\omega$. Sample values for the area of the horizon are presented in Table \ref{tab:abslimit}. In Sec.~\ref{sec:results} we compare the low-frequency limits of the absorption cross section with the numerical results for the quantity \eqref{eq:total-abs}, obtaining excellent agreement.

\subsection{High-frequency regime}
At high frequencies, under the eikonal approximation, the wave propagates along null geodesics which pass orthogonally through the initial wavefront. Hence, an analysis of absorption can be made by computing the capture cross section of null geodesics impinging on a Kerr BH from infinity. Calculations of the capture cross section may be found in Refs.~\cite{chandra,Zakharov:2005nb}. Below, we develop a complementary approach which emphasises the geometrical aspects of the calculation. 

\subsubsection{Capture cross section}

\begin{figure*}
\centering
\begin{tabular}{lcr}
 \includegraphics[height=8.2cm]{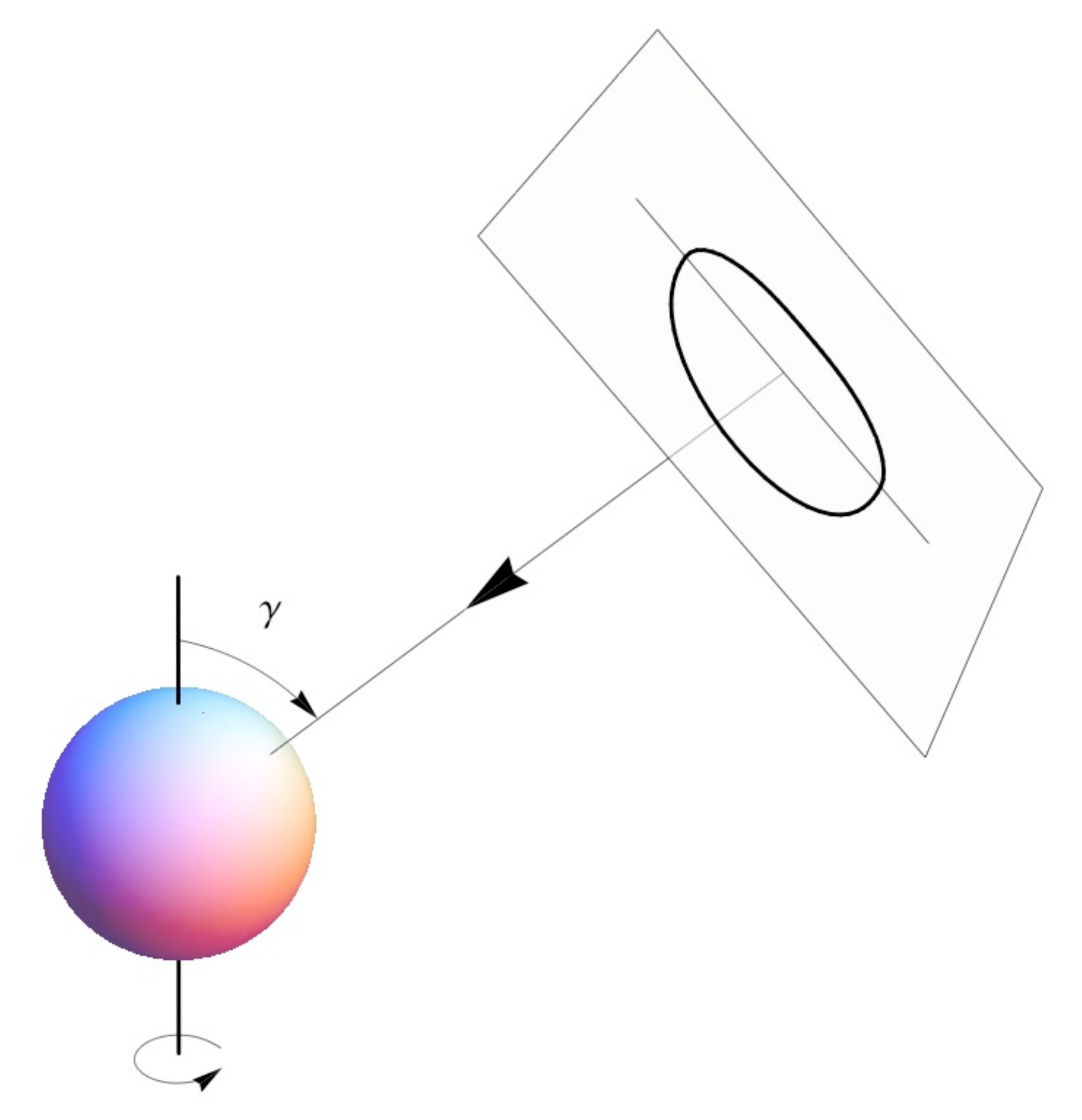} &
 \hspace{0.4cm} &
 \includegraphics[height=6cm]{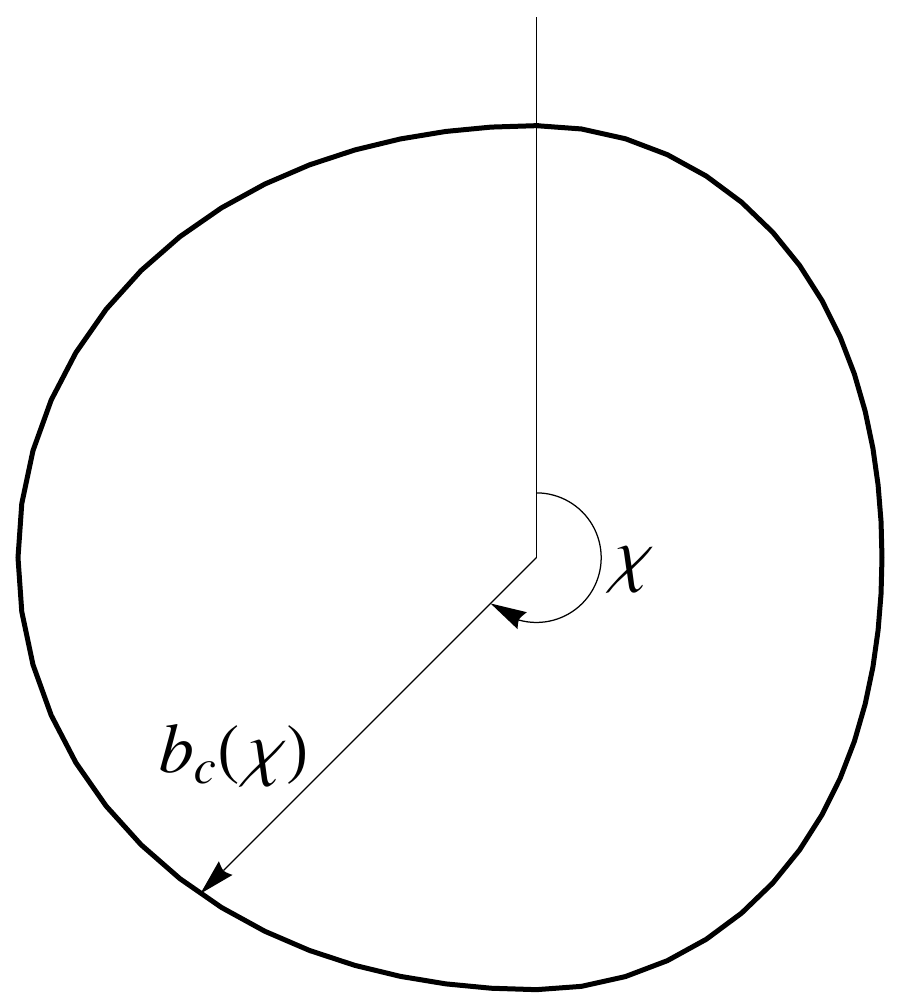}
\end{tabular}
 \caption{
  Illustrating a planar wave impinging upon a Kerr black hole. 
  The left plot shows a segment of planar wave impinging upon a black hole at angle of incidence $\gamma$ (where $\gamma$ is angle between the black-hole rotation axis and the direction of incidence). The right plot shows the locus of absorption, corresponding to that part of the wavefront which is absorbed in the geometric-optics limit. The locus is described by $b_c(\chi)$, where $\chi$ is the angle between a point on the surface and the projection of the BH rotation axis, and $b_c$ is the critical impact parameter.}
  \label{fig:setup}
\end{figure*}

Figure \ref{fig:setup} illustrates the scenario: a planar wavefront impinges upon a rotating black hole, at an angle of incidence $\gamma$. The capture cross section $\siggeo$ is the area of the `locus of absorption' which is traced on the incident planar surface. We may write
\be
\siggeo = \int_{-\pi}^{\pi} \frac{1}{2} b_c^2(\chi, \gamma) d\chi , \label{eq:capture}
\ee
where $\chi$ an angle defined on the planar surface, measured from the rotation axis in a corotating sense. Here, $b_c(\chi, \gamma)$ is the `critical' impact parameter, which corresponds to the marginal case of a null geodesic that asymptotically approaches a constant-radius photon orbit. 

\begin{figure}
 \includegraphics[width=0.9\columnwidth]{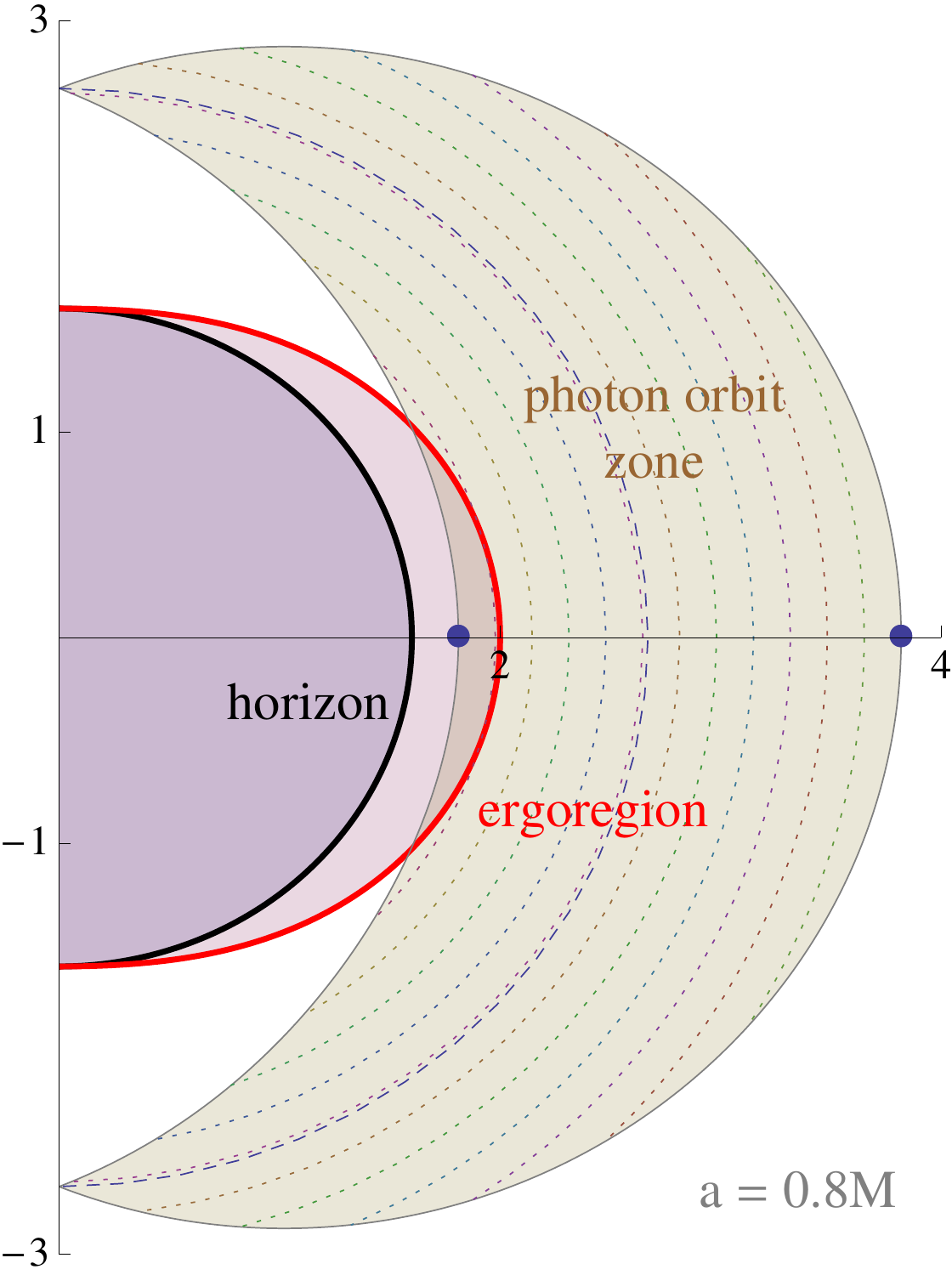}
\caption{Schematic illustration of a slice of a Kerr black hole. Here, $M=1$, $a/M=0.8$ and the event horizon [black] and ergoregion [red] are shown as solid lines. The photon orbit zone, marked in beige, is spanned by the family of constant-radius null geodesics, shown as dotted lines. Special cases include the polar orbit, which runs from pole to pole (and precesses around the black hole), shown as a dashed line at $r / M \approx 2.67$; and the co- and counterrotating equatorial orbits marked by blue dots at $r / M \approx 1.81$ and $3.82$, respectively.}
 \label{fig:schematic}
\end{figure}

For a Schwarzschild black hole, the constant-radius photon orbit occurs at $r = 3M$ (the `light-ring'), and the set of such orbits defines a surface known as the `photon sphere'. In the Kerr case, the radius of the null orbit depends on the azimuthal angular momentum. Figure \ref{fig:schematic} shows that the set of all constant-radius null orbits defines a `photon orbit zone'. Each point in this zone is associated with a constant-radius null geodesic, which is the asymptote of a null ray encroaching from spatial infinity \cite{Teo:2003}. 

To find $b_c(\chi, \gamma)$ we solve the geodesic equations which are obtained with Hamilton-Jacobi methods. Geodesics on Kerr are governed by four first-order equations, and three constants of motion: energy $E$, azimuthal angular momentum $L_z$ and Carter constant $Q$. The first step is to establish the relationship between the constants of motion and the values of $b$ and $\chi$ for the null ray passing orthogonally through the planar wavefront. Without loss of generality, let us assume that the wave is impinging along the $\phi = 0$ direction. We may introduce an `impact vector' $\mathbf{b}$ with Cartesian components $\mathbf{b} = [b \cos \gamma \cos \chi, b \sin \chi, - b\sin \gamma \sin \chi]$. This corresponds to a ray with the following constants of motion:
\begin{eqnarray}
\hat{L}_z \equiv L_z / E &=& b \sin \chi \sin \gamma, \\
\hat{Q} \equiv Q / E^2 &=& b^2 \cos^2 \chi + (b^2 \sin^2 \chi - a^2) \cos^2 \gamma .
\end{eqnarray}
The next step is to find the critical radius and impact parameter, $r_c$ and $b_c$, for the direction $\chi$, by solving
\be
R(r_c)=0, \quad \quad
\frac{\pa R(r_c)}{\pa r}=0, \label{eq:RRprime}
\ee
where
\be
R(r) = \left( (r^2 + a^2) - a \hat{L}_z \right)^2 - \Delta \left( (\hat{L}_z - a)^2 + \hat{Q} \right) .
\ee
By solving Eq.~(\ref{eq:RRprime}) we get a pair of values $(r_c(\chi,\gamma), b_c(\chi,\gamma))$, corresponding to the radius of the photon orbit and the critical impact parameter for a null ray that passes through the incident wavefront at an angle $\chi$ relative to the rotation axis, as shown in Fig.~\ref{fig:setup}. The capture cross section is computed by inserting $b_c(\chi, \gamma)$ into Eq.~(\ref{eq:capture}).

\begin{figure}
\includegraphics[width=0.95\columnwidth]{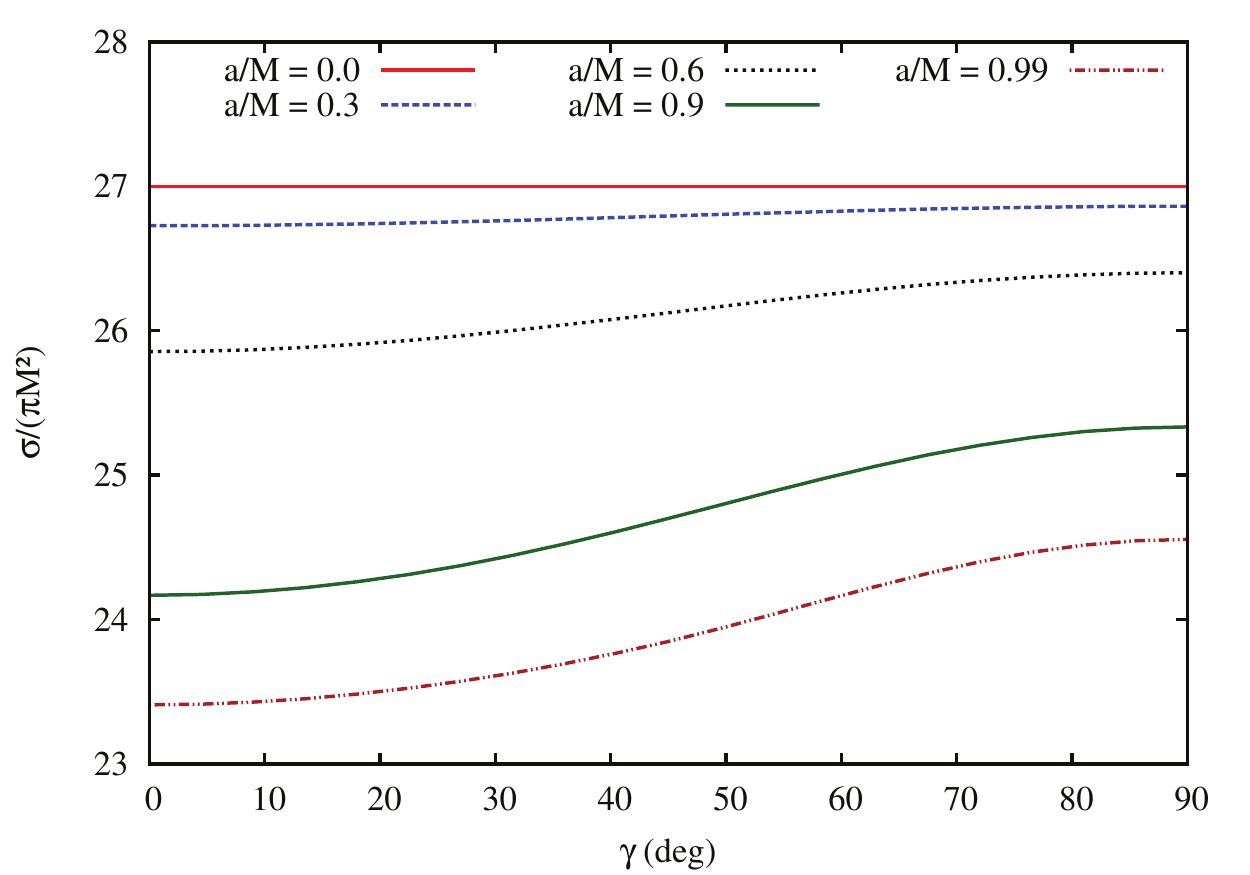}
\caption{Geodesic capture cross section $\sigma$ as a function of angle of incidence $\gamma$, where $\gamma = 0$ corresponds to incidence along the black hole's axis of rotation (and $\sigma$ is symmetric under $\gamma \rightarrow \pi - \gamma$). 
The geodesic capture cross section is the high-frequency asymptote for the planar wave absorption cross section. 
}%
\label{fig:capture-csec}%
\end{figure}

Figure \ref{fig:capture-csec} shows the geodesic capture cross section as a function of angle of incidence $\gamma$ for a variety of black hole spins $a$. 
Some values for the capture cross section in the special case of on-axis incidence ($\gamma = 0$) are presented below in Table \ref{tab:abslimit}. These values will be compared with the numerical results for the absorption cross section exhibited in Sec.~\ref{sec:results}.

\begin{table}[h]
\begin{tabular}{l c c c c c}
\hline\hline
$a~[M]$ & $0.00$ & $0.30 $  & $0.60$& $0.90$ & $0.99$\\\hline
$\sigma(\omega\approx 0)~[\pi M^2]$ & $16.000$ & $15.631$ &$14.400$& $11.487$ & $9.128$\\\hline
$\sigma(\omega\gg M)~[\pi M^2]$ & $27.000$ & $26.726$ & $25.855$ & $24.168$ & $23.409$\\	\hline\hline
\end{tabular}
\caption{Low- and high-frequency limits of the absorption cross section for the different choices of $a$ exhibited in the plots of the Sec.~\ref{sec:results}. The high-frequency results presented here are for on-axis incident null geodesics.}
\label{tab:abslimit}
\end{table}



\subsubsection{Sinc approximation\label{subsec:sinc}}

In the 1970s, Sanchez \cite{Sanchez:1977si} found that, at high frequencies, the absorption cross section for a Schwarzschild BH oscillates around the geometric capture cross section with a peak-to-peak interval of $\Delta \omega = 1/\sqrt{27}M$. In one sense, oscillations arise from the contributions of successive partial waves to Eq.~(\ref{eq:total-abs}). In a complementary sense, as recently shown in Refs.~\cite{DEFF, DFR} using complex angular momentum (CAM) methods, the oscillations are related to the properties of the unstable photon orbit. For a scalar field absorbed by a spherically-symmetric BH, it was shown that \cite{DEFF}
\beq
\sigma / \sigma_\text{geo} \sim  1 - 8 \pi \beta \, e^{-\pi \beta} \, \text{sinc} \left( 2 \pi \omega / \Omega \right){,}
\label{eq:sigma-CAM1}
\eeq
where $\beta \equiv \Lambda / \Omega$, with $\Omega$ the frequency of the null orbit, and $\Lambda$ the associated Lyapunov exponent. In the Schwarzschild case, $\sigma_\text{geo} = \pi b_c^2$ is the geodesic capture cross section and $\Omega = 1 / (\sqrt{27} M) = 1/b_c$ with $\beta = 1$. The oscillatory term arises from a (high-frequency approximation to a) sum over Regge poles. Regge poles are characteristic resonances of the spacetime closely related to the quasinormal modes \cite{Dolan:2010, Yang:2012, Yang:2013}. The idea that oscillations in absorption cross sections provide information about the properties of the null orbits is an intriguing one, which surely deserves further investigation in non-spherically-symmetric cases, such as Kerr.

In Sec.~\ref{sec:results} we show that oscillations around the capture cross section are also present in the Kerr context, and, for general angles of incidence, these oscillations exhibit a richer spectrum. The oscillations arise from the superposition of partial contributions which now depend on azimuthal number $m$ as well as on $l$. From the complementary viewpoint, these oscillations are related to the spectrum of Regge poles, which also depend on both $l$ and $m$.

\begin{figure}
\includegraphics[width=0.95\columnwidth]{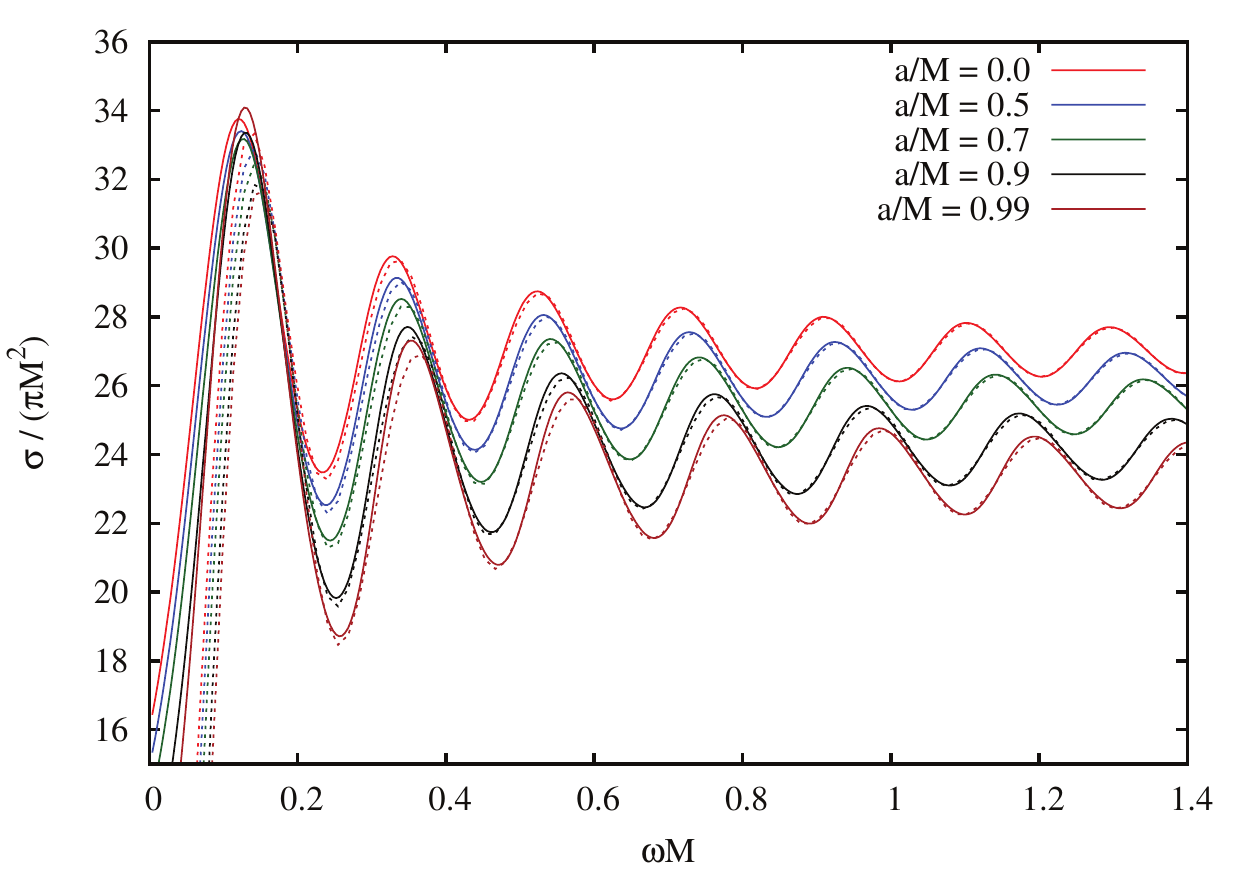}
\caption{The sinc approximation for on-axis incidence ($\gamma = 0$). The solid lines show the numerically-determined cross section (cf.~Fig.~\ref{fig:on-axis}), and the dashed lines show the sinc approximation, Eq.~(\ref{eq:sigma-CAM2}), for a range of $a/M$ and $\omega M$.
}%
\label{fig:sinc} 
\end{figure}

In the special case in which the plane wave is incident along the axis of rotation ($\gamma = 0$ or $\pi$), a slightly-modified version of  Eq.~(\ref{eq:sigma-CAM1}) is still valid, even though the BH itself is not spherically symmetric. One subtlety is that we need to take account of the spheroidal harmonics in Eq.~(\ref{absorption}). Progress can be made with an asymptotic relation, obtained using the WKB techniques of Ref.~\cite{Yang:2012}:
\beq
|{S}_{\omega l 0}(\gamma=0)|^2 = \frac{1}{4\pi^2} \frac{\partial A}{\partial L} {.}
\eeq
Here $L = l+1/2$ and $A$ is the angular eigenvalue of the spheroidal equation for $m=0$. We may then use the following expansion
\beq
\frac{A}{L^2} = 1 - \frac{1}{2} \alpha^2 + \frac{1}{32} \alpha^4 + \frac{5}{8192} \alpha^8 + \mathcal{O}(\alpha^{12}) ,
\eeq
which is valid in the regime $\alpha \equiv a \omega / L < 1$. It follows that $2 \pi^2 |{S}_{\omega l 0}(\gamma=0)|^2 = L {-} \frac{\alpha^4 L}{32} + \ldots$, and the subdominant term is very small near the Regge pole, and may be neglected. The other steps in the derivation of Ref.~\cite{DEFF} follow through unchanged, and we arrive at
\beq
\sigma / \sigma_\text{geo}  \sim 1 -  \frac{8 \pi \beta  \, e^{-\pi \beta}}{\Omega^2 b_c^2} \, \text{sinc} \left( 2 \pi \omega / \Omega \right) ,
\label{eq:sigma-CAM2}
\eeq 
where again $\beta = \Lambda / \Omega$. In the on-axis case, $b_c$, $\Omega$ and $\Lambda$ can be written in closed form; the relevant expressions are found in Eqs.~(18), (22) and (24), respectively, of Ref.~\cite{Dolan:2010}. Note that now $b_c \neq 1 / \Omega$, for $a \neq 0$. In Fig.~\ref{fig:sinc} we plot a selection of results obtained through Eq.~(\ref{eq:sigma-CAM2}), and compare with numerically-determined cross sections.

\subsubsection{Semi-analytic approximation\label{subsec:semianalytic}}
In the high frequency regime, the behaviour of the transmission factors are closely linked to the properties of null orbits, via 
\beq
\Gamma_{\omega l m} \sim \left[ 1 + \exp \left(-2 \pi (\omega - \Omega L) / \Lambda \right) \right]^{-1}
\label{eq:transapprox}
\eeq
(cf.~Eq.~(15) in Ref.~\cite{DEFF}). Here $\Omega(a/M, \mu)$ and $\Lambda(a/M, \mu)$ are, respectively, the orbital frequency and Lyapunov exponent associated with a null orbit with angular momentum ratio $\mu \equiv m / L$. Accurate semi-analytic approximations for $\Omega$ and $\Lambda$ are given in Ref.~\cite{Yang:2012}: see Eq.~(2.35), (2.36) and Eq.~(2.40).

Figure \ref{fig:transapprox} shows the transmission factors $\Gamma_{\omega l m}$ as a function of frequency $M\omega$, for the case $l=5$, $a = 0.9M$, and $-l \le m \le l$. It shows that Eq.~(\ref{eq:transapprox}) provides an excellent approximation for estimating the transmission factors. 

\section{Numerical results}
\label{sec:results}

\begin{figure}
\includegraphics[width=0.95\columnwidth]{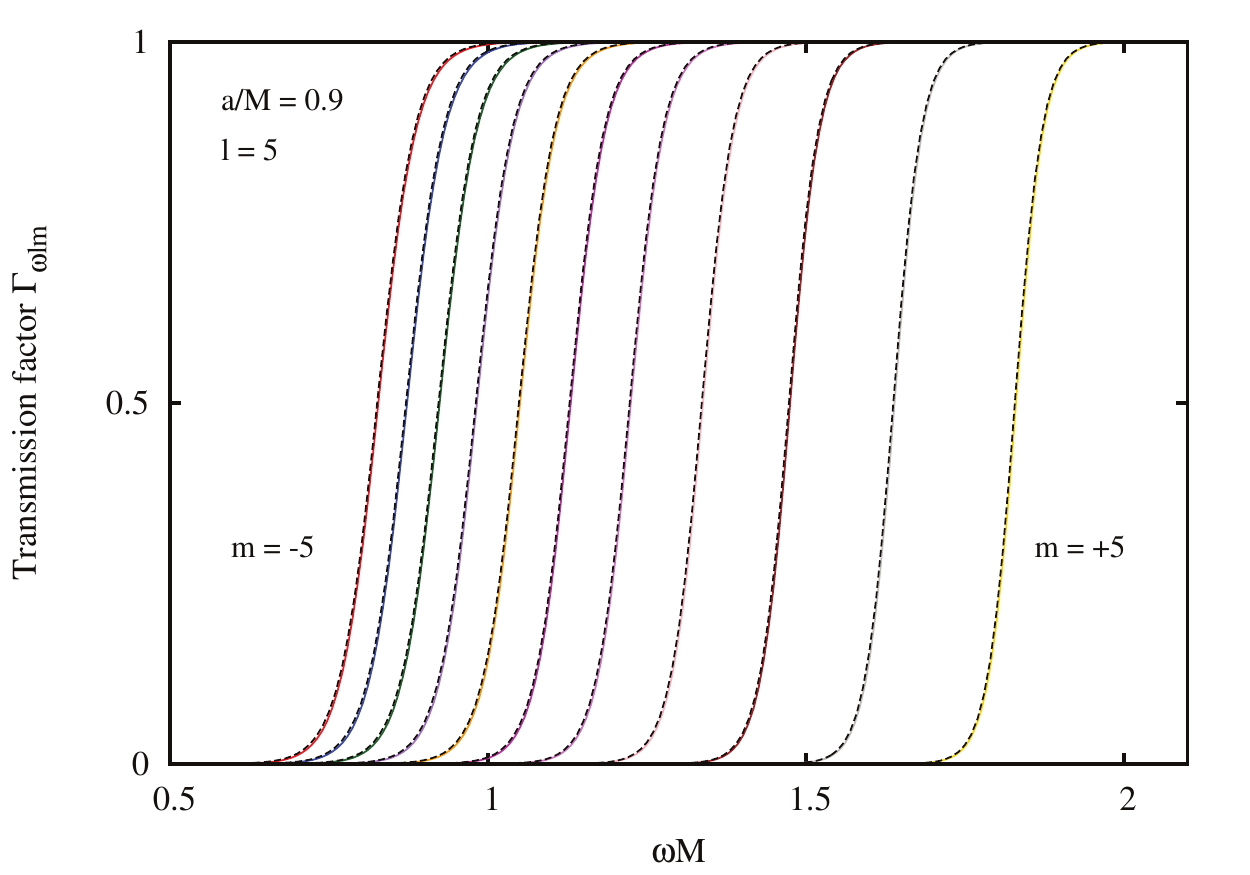}
\caption{
Transmission factors $\Gamma_{\omega l m}$ for angular multipoles $l = 5$ and $-l \le m \le l$. The solid lines show numerical solutions of Eq.~(\ref{eq:trans}), and the dashed lines show the semi-analytic approximation, Eq.~(\ref{eq:transapprox}). 
}%
\label{fig:transapprox} 
\end{figure}

%
%
\begin{figure}[h]
\includegraphics[width=0.95\columnwidth]{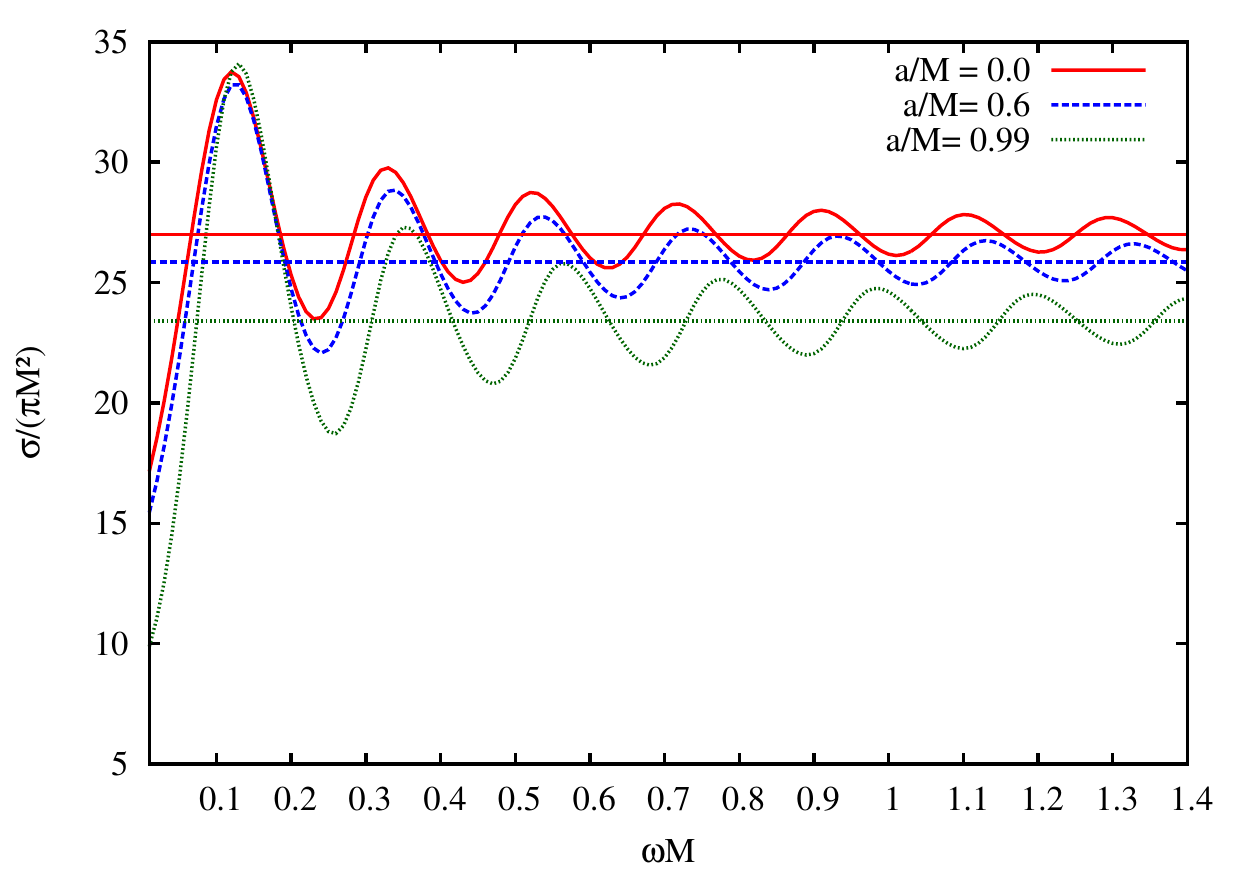}
\caption{On-axis ($\gamma = 0$) absorption cross section for $a/M=0.00,\,0.60$ and $0.99$. The horizontal lines represent the high-frequency limits.  We see that the general pattern of the on-axis absorption cross section, even for rapidly rotating BHs, is similar to the case of spherical ($a=0$) BHs.}%
\label{fig:on-axis}%
\end{figure}
\begin{figure*}[h]
\includegraphics[width=0.95\columnwidth]{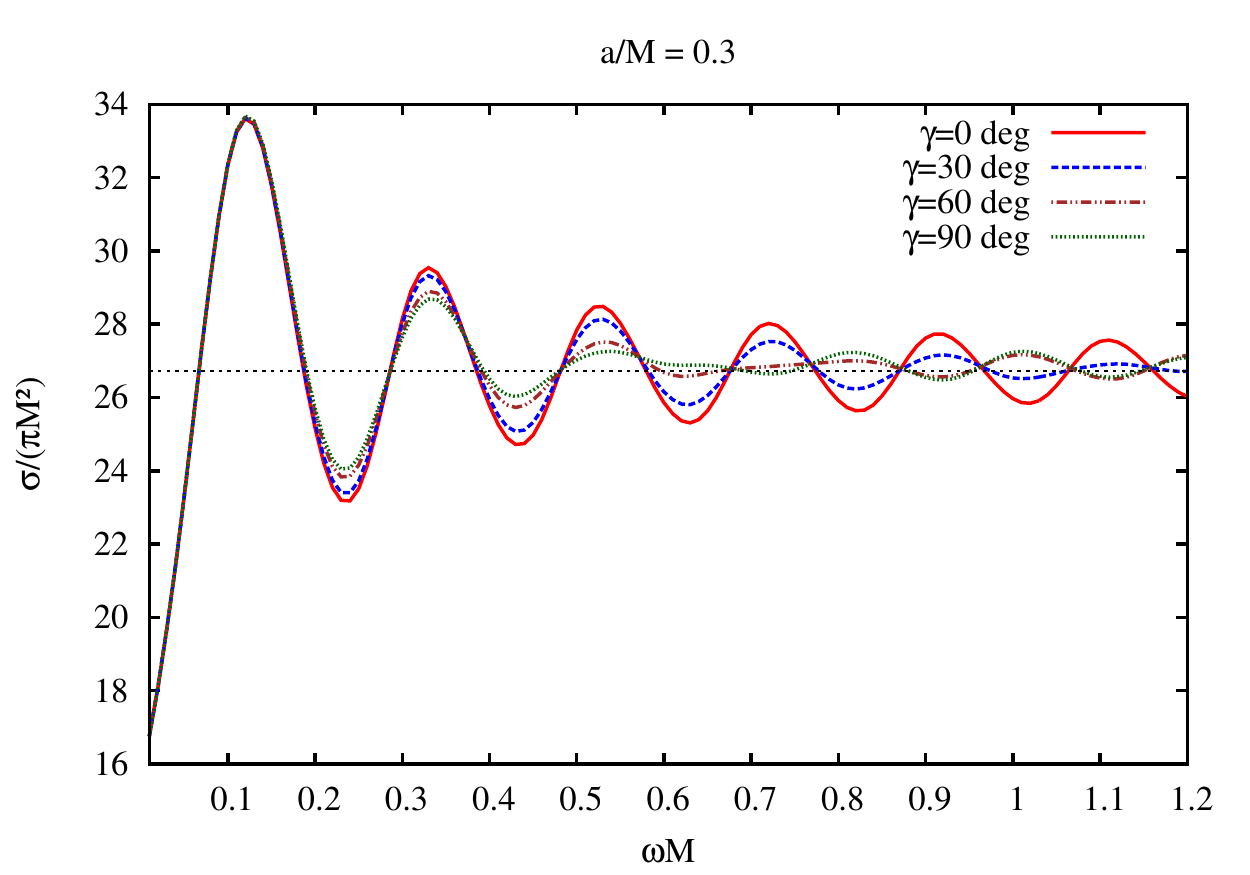}\includegraphics[width=0.95\columnwidth]{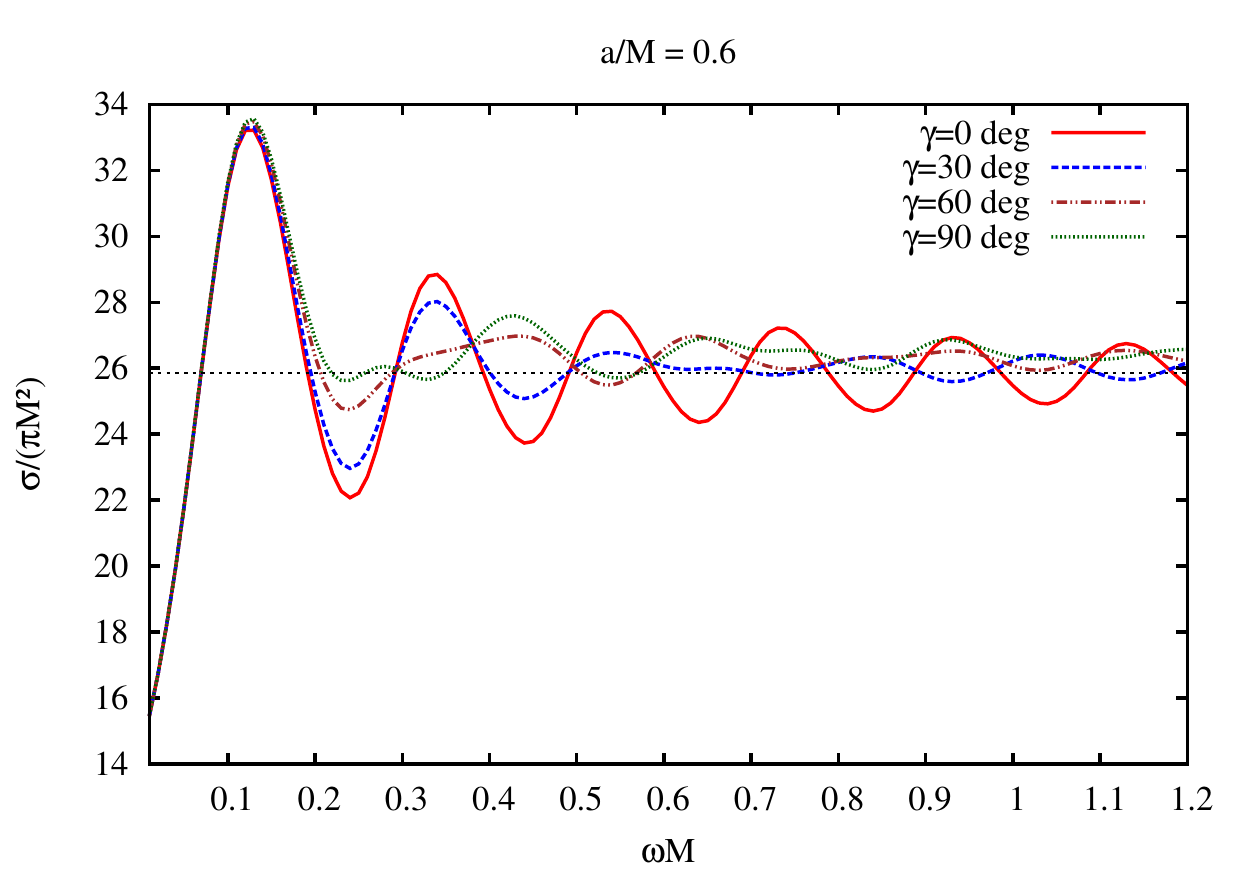}
\includegraphics[width=0.95\columnwidth]{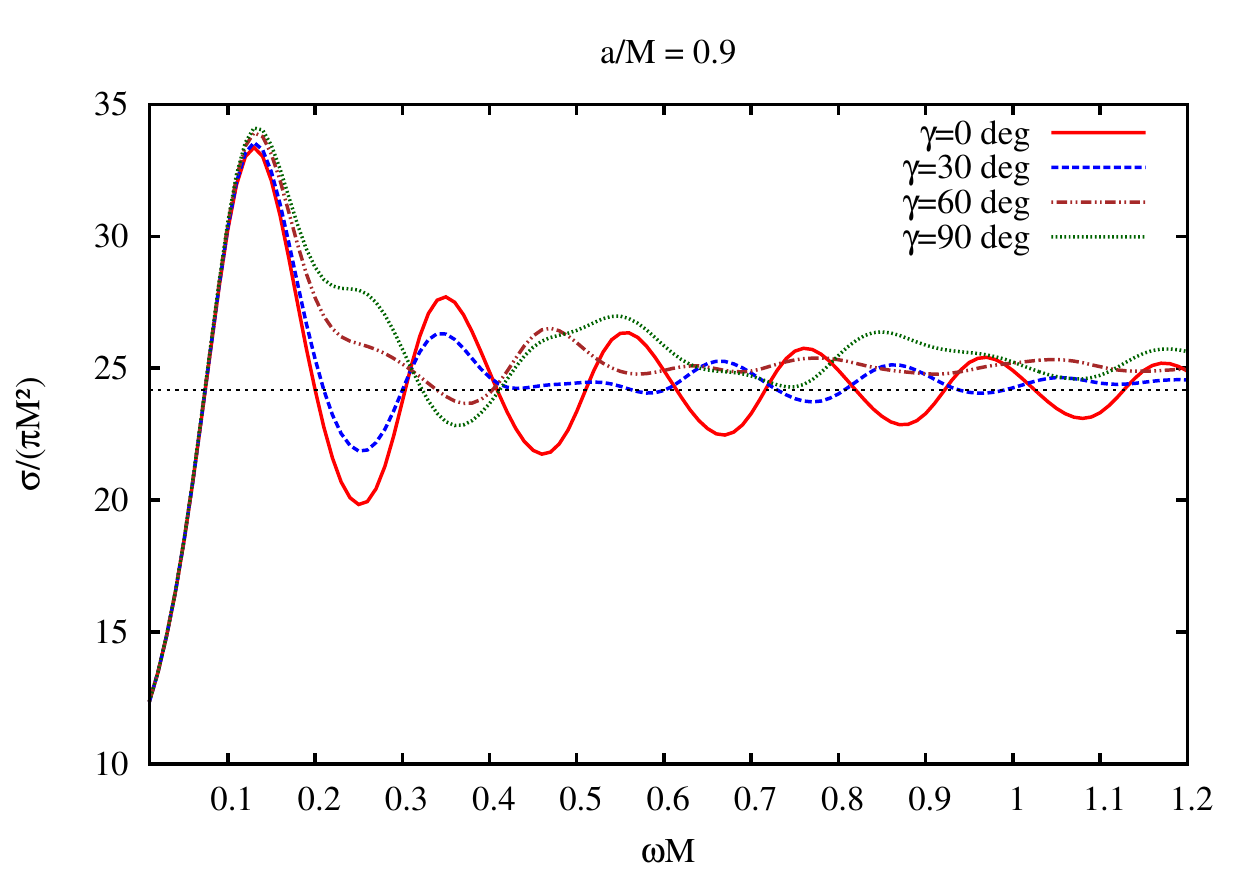}\includegraphics[width=0.95\columnwidth]{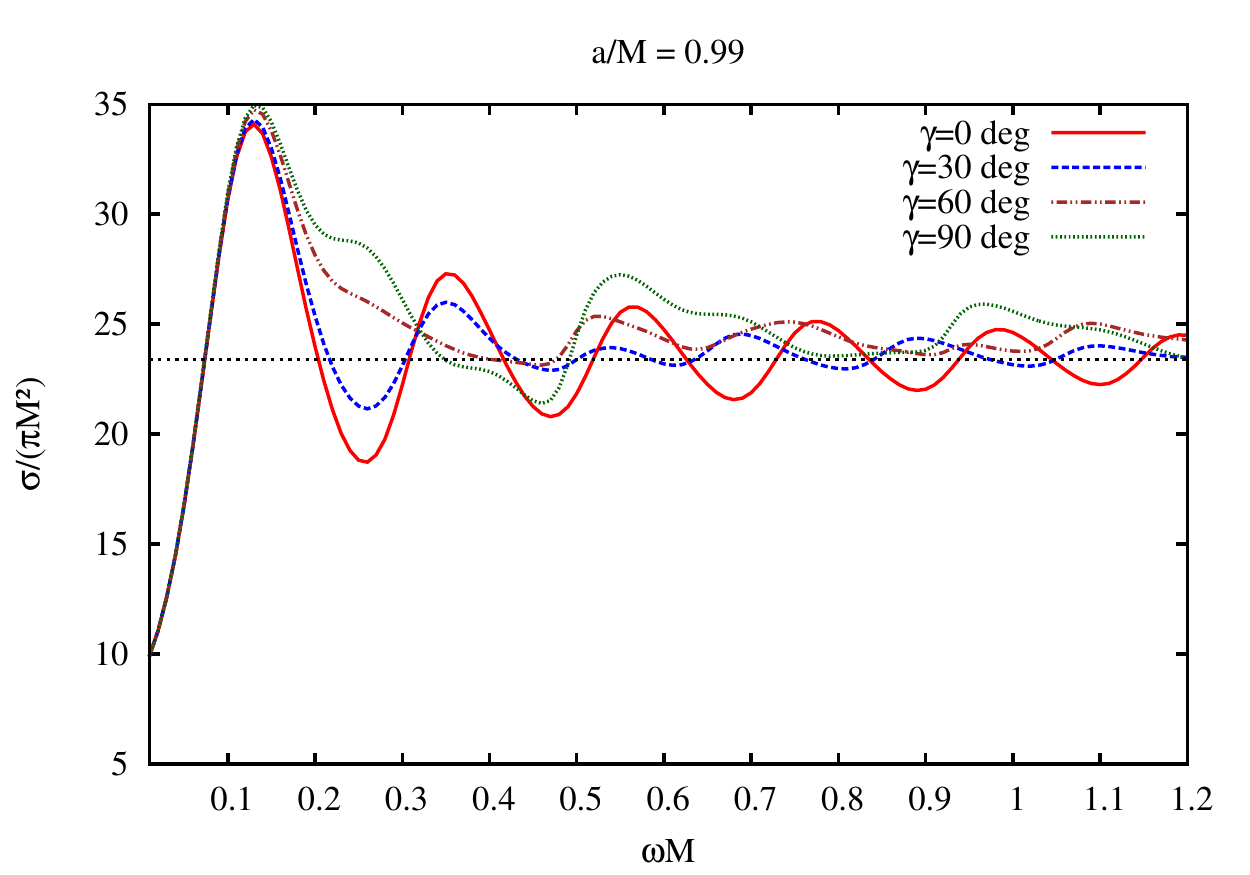}
\caption{Off-axis absorption cross section for $a/M=0.30,~0.60,~0.90$ and $0.99$. For comparison, we also exhibit the on-axis case ($\gamma=0$), and its high-frequency limit (horizontal lines). We see that the oscillation pattern for the off-axis cases differs considerably from the regular one exhibited in the on-axis case.}%
\label{fig:off-axis}%
\end{figure*}

%

\begin{figure*}[h]
\includegraphics[width=0.95\columnwidth]{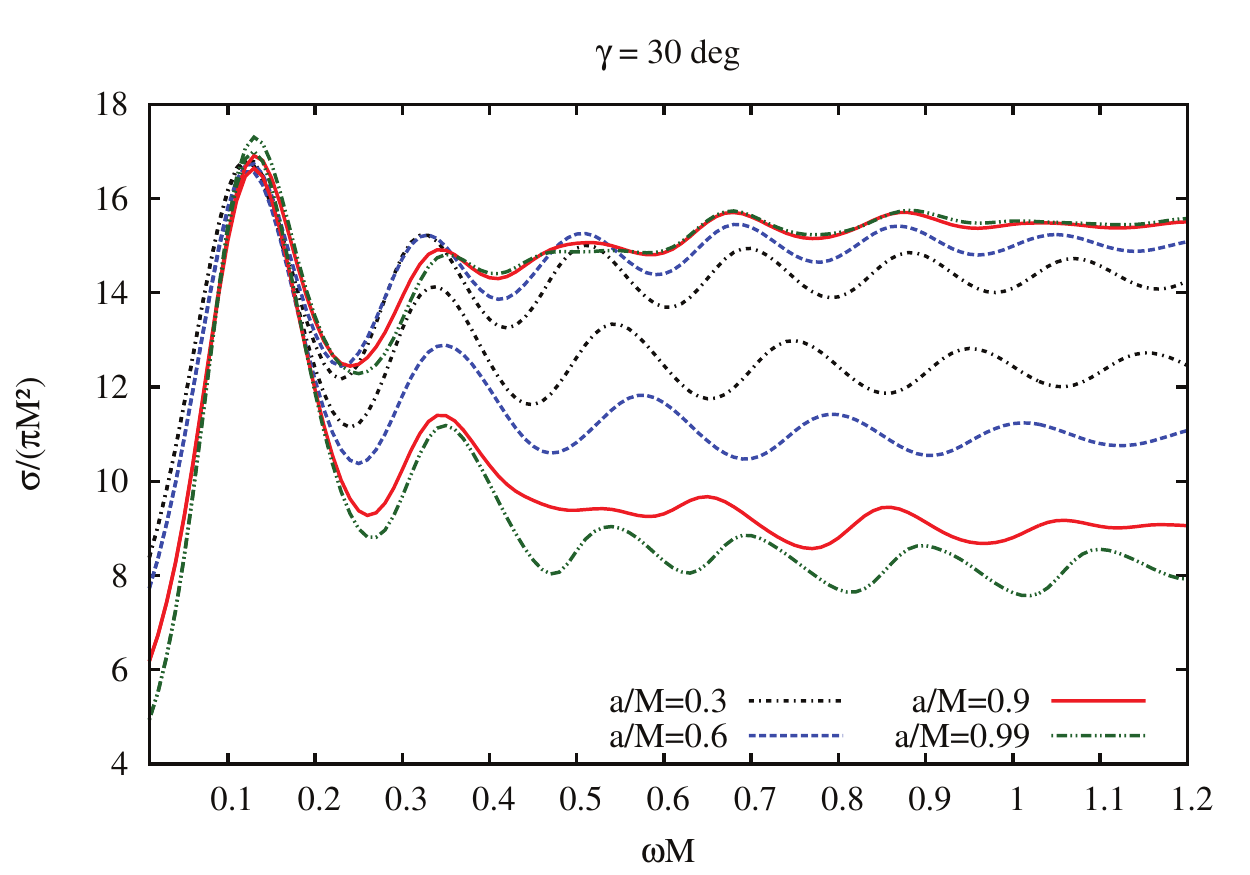}\includegraphics[width=0.95\columnwidth]{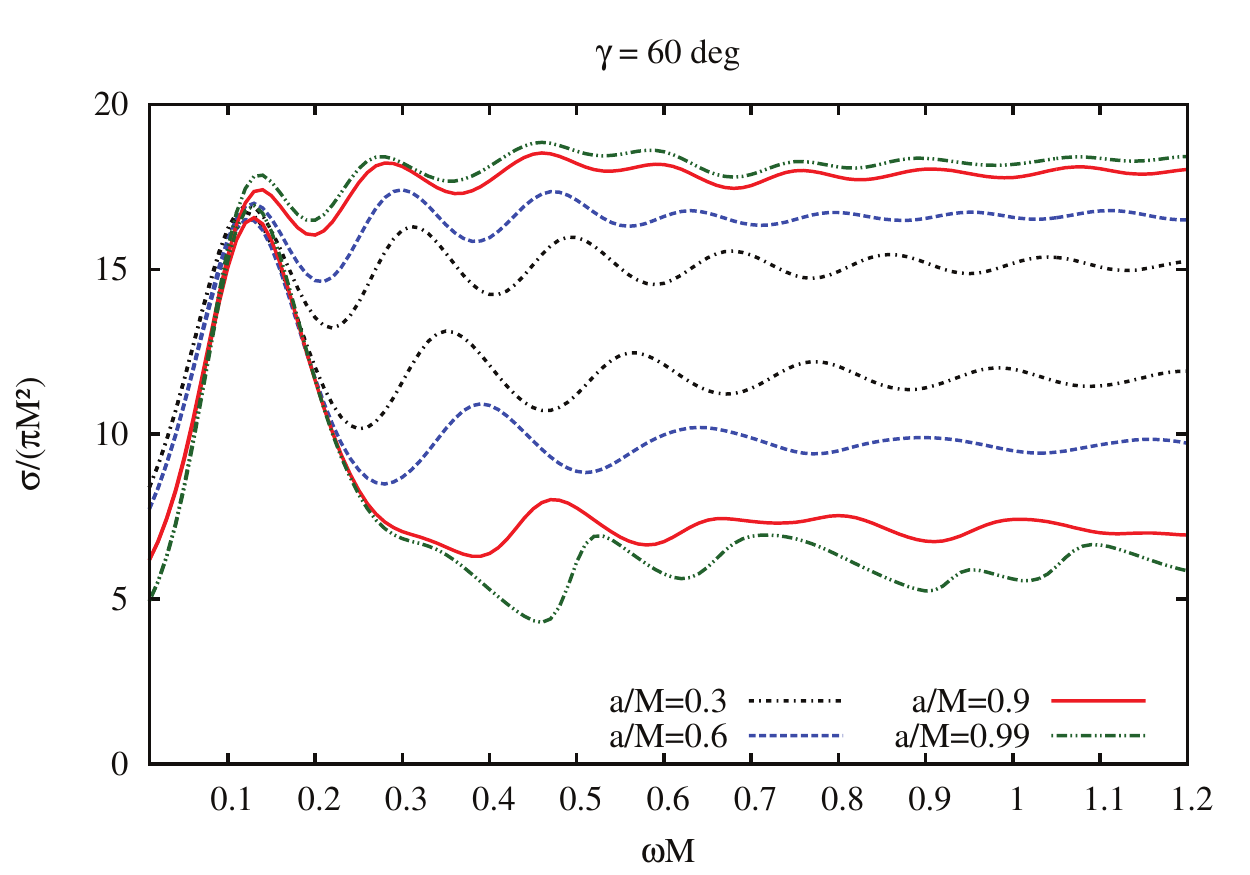}
\includegraphics[width=0.95\columnwidth]{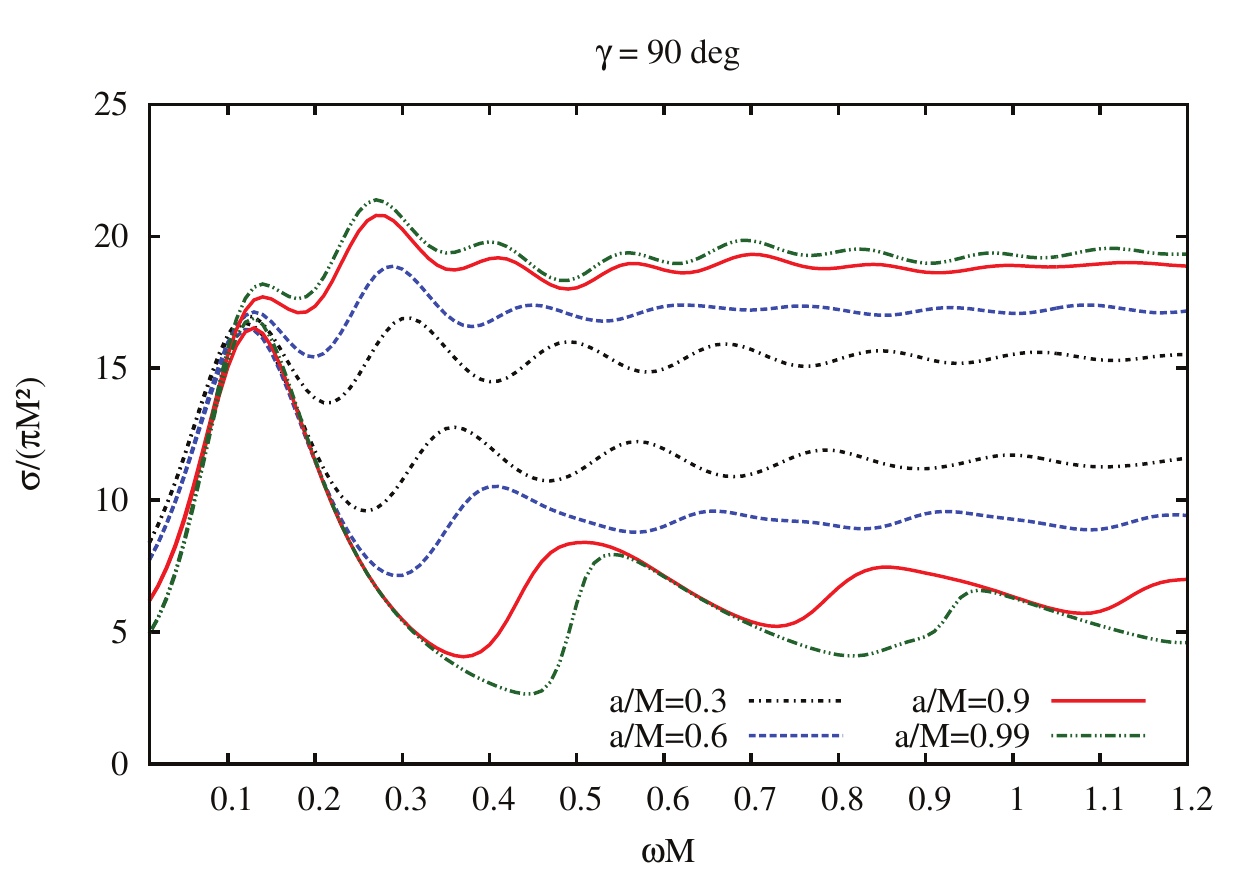}
\caption{Co- ($\sigma^+$) and counterrotating ($\sigma^-$) absorption cross section for $\gamma=30,\,60$, and $90$ degrees. The curves are plotted for different BH rotation parameters, namely $a/M=0.30,\,0.60,\,0.90$, and $0.99$. The counterrotating absorption cross sections are larger than the correspondent corotating ones, and their separation becomes larger as $a$ increases.}%
\label{fig:co-counter}
\end{figure*}

\begin{figure*}[h]
\includegraphics[width=0.95\columnwidth]{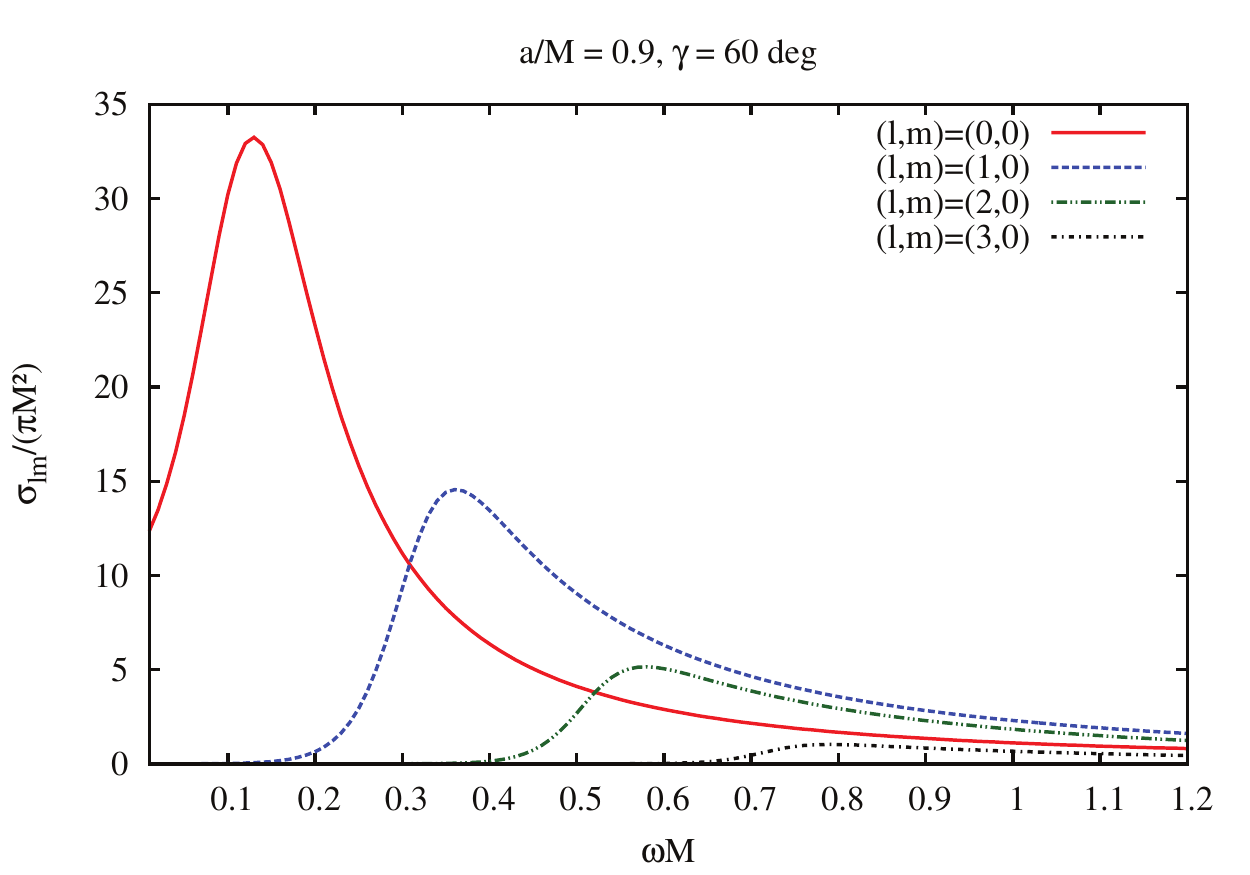}\includegraphics[width=0.95\columnwidth]{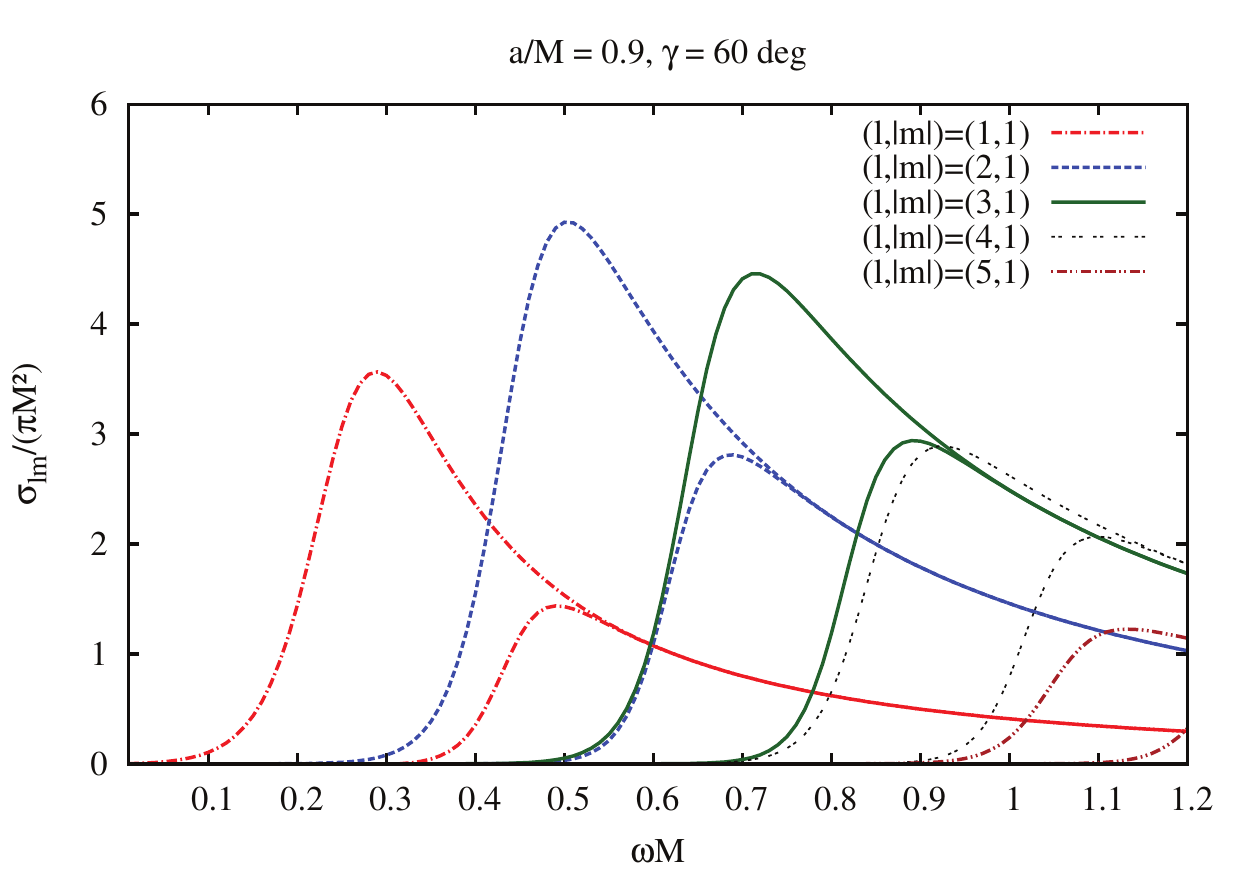}
\includegraphics[width=0.95\columnwidth]{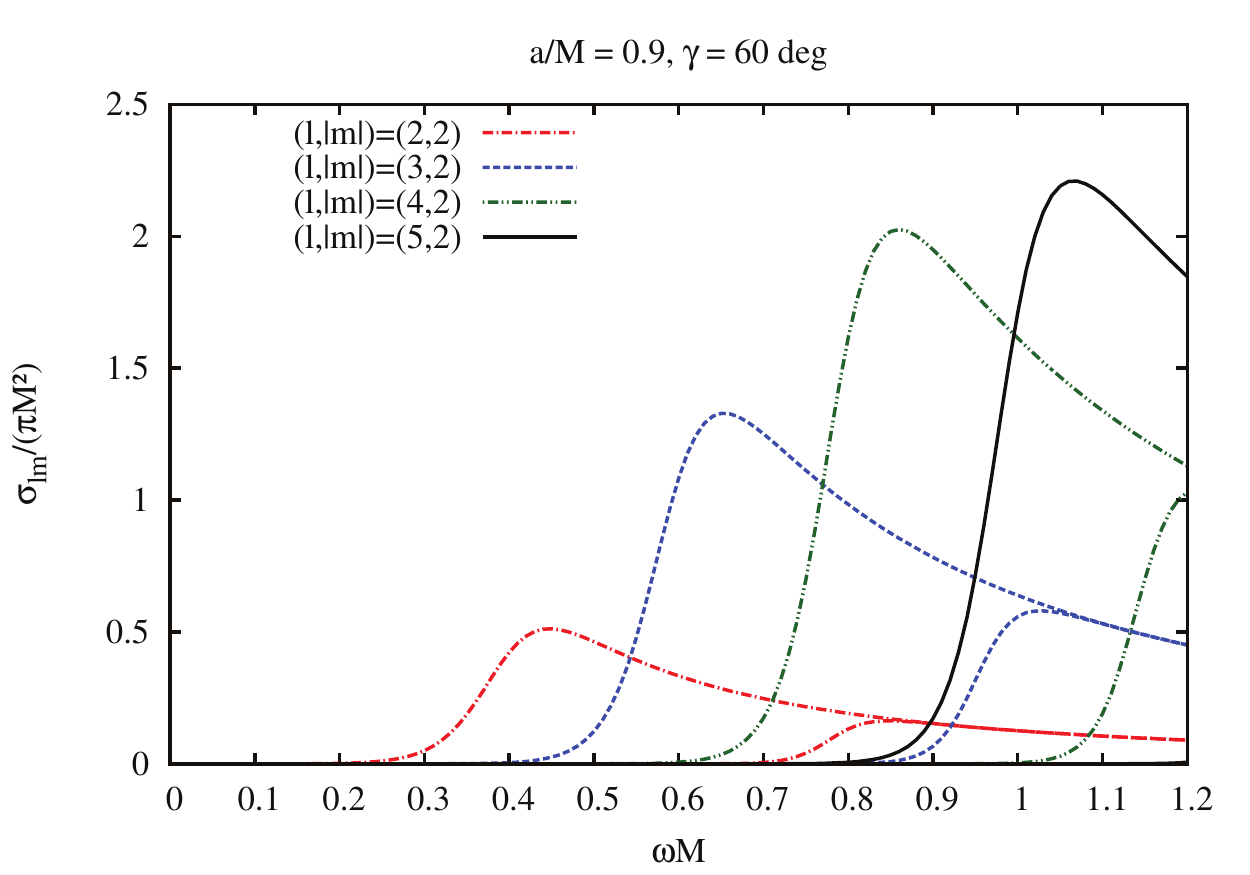}\includegraphics[width=0.95\columnwidth]{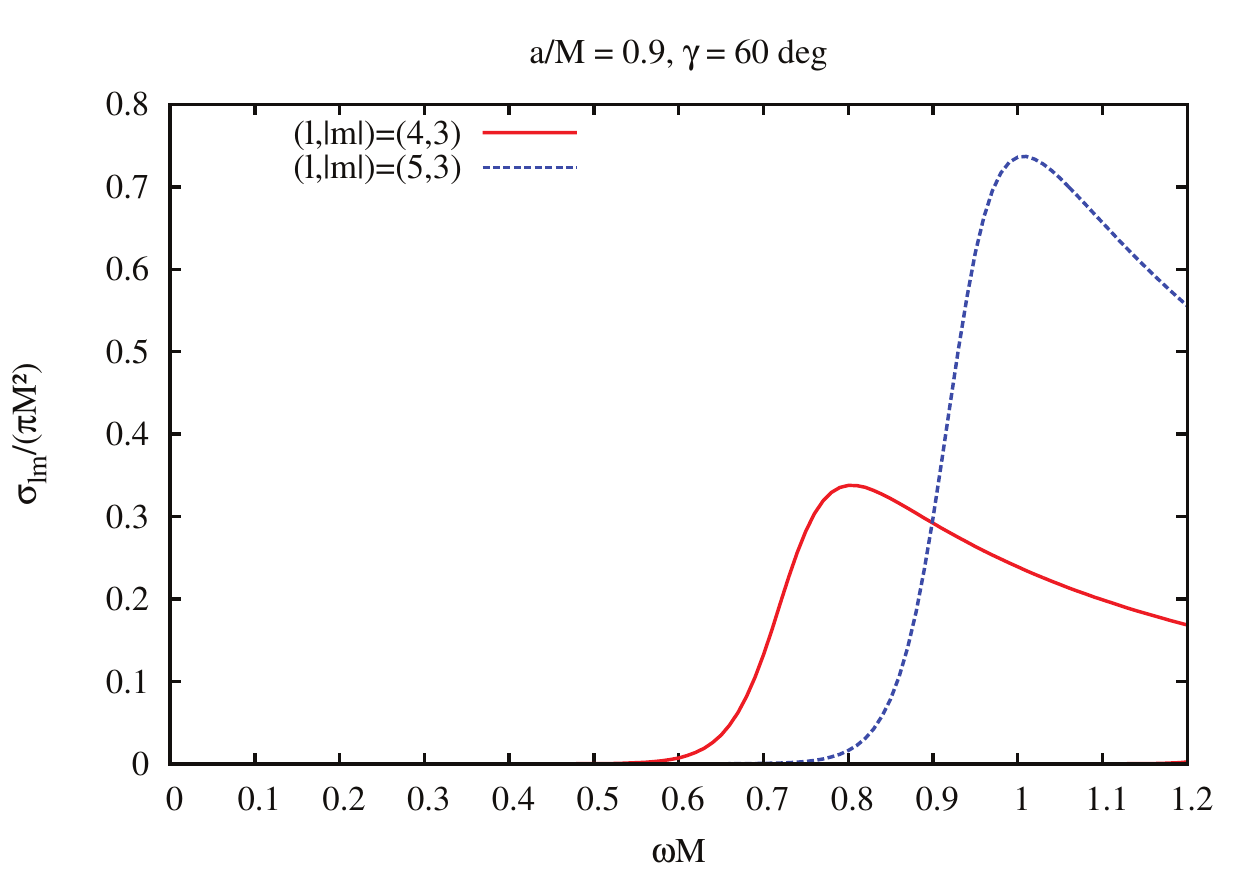}
\label{fig:partial}
\caption{Partial absorption cross sections for $a/M=0.90$ and $\gamma=60$ degrees. We plot the main contributions for fixed values of $l$ and $m$, both for corotating (curves with lower peaks) and counterrotating (higher peaks) cases.}%
\label{fig:off-axis-partial}%
\end{figure*}
\begin{figure*}[h]
\includegraphics[width=0.95\columnwidth]{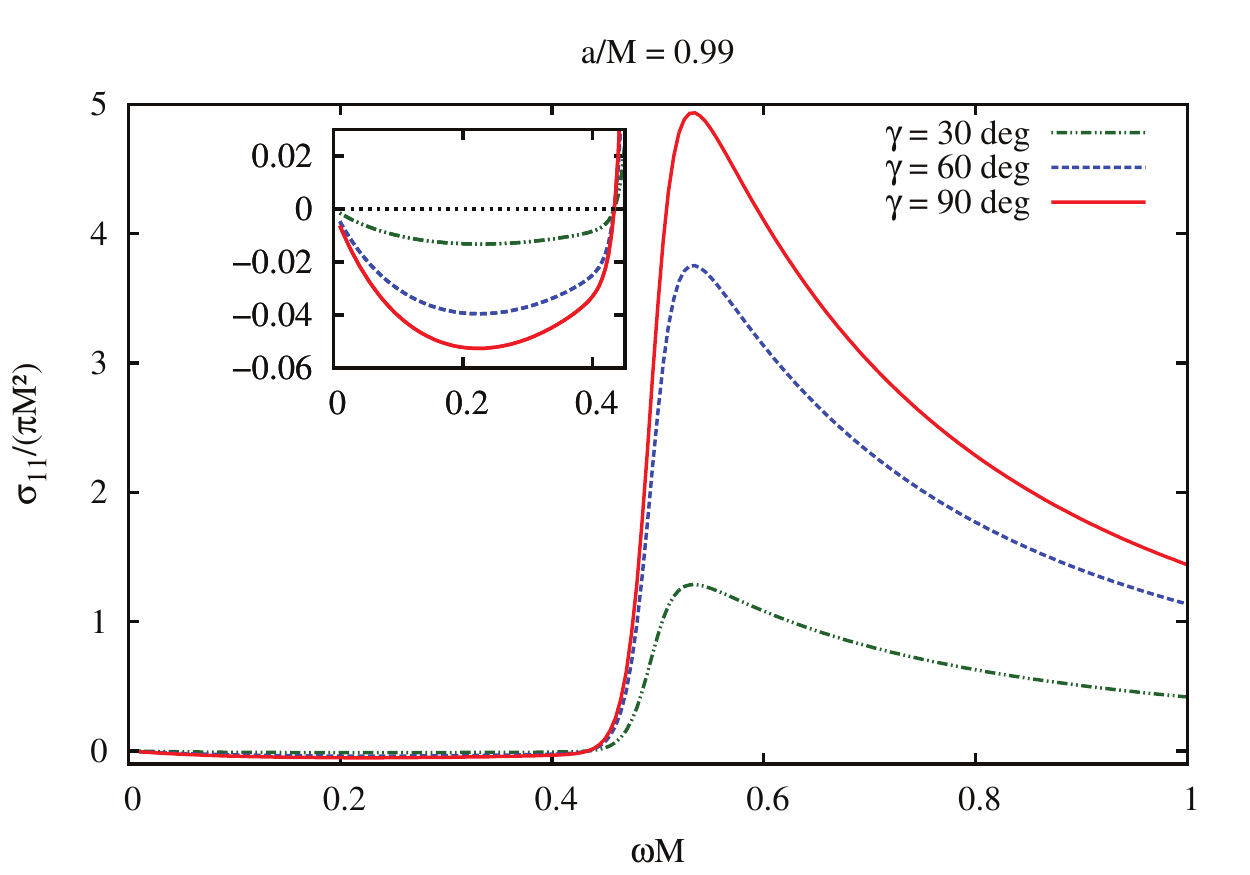}
\caption{Zoom at the mode with $l=m=1$ for $\gamma=30,\, 60$, and $90$ degrees, with $a/M=0.99$. Superradiance is more evident for this mode, although being very small even in this case.}%
\label{fig:zoom-abs}%
\end{figure*}%


In this section we present our numerical results for the scalar absorption cross section of the Kerr BH. We give particular attention to the off-axis absortion cross section, i.e., the $\gamma  \neq 0$ cases, which exhibit many distinct features when compared to the spherically symmetric case.

As for the numerical precision, we have considered the summation in Eqs.~\eqref{eq:expansion1} and \eqref{eq:expansion2} until the next term contributes less than $10^{-3}$ of the total value. This makes the computation more efficient, as we do not need to integrate to very large radii in order to obtain convergent results. In our case, the maximum radius $r_\infty$ is typically $r_\infty/r_+\sim 10^2$ and the numerical outer horizon $r_h$ is such that $(r_h/r_+-1) \sim 10^{-2}$.

The numerical upper limit in the $l$ summation in Eq.~\eqref{eq:total-abs}, $l_{\text{max}}$, should be considered carefully, in order to properly compute the total absorption cross section. The convergence of Eq. \eqref{eq:total-abs} depends strongly on the value of the wave frequency $\omega$. For higher values of $\omega$ one should take higher values of $l_{\text{max}}$. For the results presented here, which are in the frequency range $0<\omega M<1.4$, we performed the summation until $l_{\text{max}}=8$. Additional terms coming from $l>8$ would contribute less than $10^{-6}$ of the total value, being unnoticeable in the data plots presented here. Our results were checked using independent codes, which increases their reliability.

In Fig.~\ref{fig:on-axis} we show the total on-axis absorption cross section for  $a/M=0.00$ (Schwarzschild case), $0.60$, and $0.99$. In the on-axis case, as the frequency is increased, the absorption cross section increases from the area of the event horizon and then oscillates regularly around the high-frequency limit given by the capture cross section of null geodesics (see Fig.~\ref{fig:capture-csec}), which is represented by horizontal lines in Fig.~\ref{fig:on-axis}.

In Fig.~\ref{fig:sinc} we compare the numerically-determined cross section with the `sinc approximation' of Eq.~(\ref{eq:sigma-CAM2}). We see that the agreement is excellent in the moderate-to-large $\omega M$ regime, which confirms the validity of Eq.~(\ref{eq:sigma-CAM2}).

In Fig.~\ref{fig:off-axis} we show the absorption cross section for a range of rotation parameters ($a/M=0.30,\,0.60,\,0.90,$ and $0.99$) and incidence angles ($\gamma=0,\,30,\,60,$ and $90$ degrees). We see that, as we move away from the on-axis case ($\gamma=0$), by increasing the incidence angle $\gamma$, and increasing the rotating parameter $a$, the absorption cross section starts to differ considerably from the regular behavior shown in Fig.~\ref{fig:on-axis}. In the high-frequency regime, $\sigma$ oscillates in an irregular way around the geodesic capture cross section. This irregular oscillatory behavior arises as a consequence of breaking the azimuthal degeneracy, so that the transmission factor becomes strongly dependent on $m$, as shown in Fig.~\ref{fig:transapprox}. In other words, there is a coupling between the BH rotation and the azimuthal number $m$, which may be interpreted as the result of frame-dragging \cite{Glampedakis:2001cx}.

The azimuthal number $m$ may be positive, which corresponds to corotating modes, or negative, which corresponds to counterrotating modes. In order to see their contribution separately, we computed the absorption cross sections, $\sigma^+$ and $\sigma^-$, as defined in Eq.~\eqref{eq:co-counter}. The results are shown in Fig.~\ref{fig:co-counter}. When we split the absorption cross section into co- ($\sigma^+$) and counterrotating ($\sigma^-$) contributions, we see that the oscillating pattern becomes more regular. Furthermore, we see that the counterrotating contributions for the total absorption cross section are larger than the corotating ones. This agrees with the null geodesic analysis, where the critical radius for retrograde orbits is larger than that for prograde orbits \cite{chandra}. We note that $\sigma^+$ and $\sigma^-$ move further apart as the rotation rate increases. The difference between the co- and counterrotating absorption cross sections is more pronounced for $\gamma=90$ deg, as a consequence of the increased importance of frame-dragging in the equatorial plane.

In Fig.~\ref{fig:off-axis-partial} we show the main partial contributions for the total absorption cross section for fixed values of $|m|$, varying $l$, according to Eq.~\eqref{absorption}. We see from Fig.~\ref{fig:off-axis-partial} that corotating ($m > 0$) and counterrotating ($m < 0$) contributions to the partial absorption cross section with the same value of $|m|$ become equal after a certain value of the frequency. This occurs when both partial waves are completely absorbed, i.e.~$\l|\olm{\mathcal{R}}^{in}/\olm{\mathcal{A}}^{in}\r|^2 = 0$, and the sign of $m$ in Eq.~\eqref{absorption} becomes irrelevant. The approximate value of the frequency at which absorption becomes significant is determined by $\Omega L$, where $\Omega$ is the frequency of the corresponding null orbit, which depends on both $m / L$ and $a / M$ (see Sec.~\ref{subsec:semianalytic}). 


Due to superradiance \cite{Press:1972zz}, the reflection coefficient can actually exceed unity for some values of $(\omega, m)$. See, for instance, Ref.~\cite{Macedo:2012zz}, where the reflection coefficient is computed for different values of $a$. For these values, as can be seen in Fig.~\ref{fig:zoom-abs}, the transmission factor and partial absorption cross section are negative, although the total absorption cross section remains positive. We recall that there is no superradiance for $m=0$, and that it is most evident for the $l=m=1$ mode.

\section{Conclusion}
\label{sec:conclusion}
We have numerically computed the absorption cross section of plane massless scalar waves incident upon Kerr BHs, for general angles of incidence, revealing the effect of black hole rotation.  In the special case of on-axis incidence, we showed that the absorption cross sections are well-described by a simple `sinc' approximation. Our result was obtained by extending the complex angular momentum method of Ref.~\cite{DEFF}. In the general case of arbitrary incidence, we showed that the absorption cross section of a Kerr BH exhibits an irregular oscillation pattern, which is in contrast to the regular oscillations shown by a Schwarzschild BH. We have taken steps to explain this effect in terms of the coupling between the azimuthal angular momentum of the field and the angular momentum of the BH. In Sec.~\ref{subsec:semianalytic} we gave a semi-analytic approximation to relate the transmission of partial waves to the properties of the null photon orbits. To explore the coupling, we have compared the corotating ($m > 0$) and counterrotating ($m < 0$) contributions to the absorption cross section. We have shown that, due to superradiance in the Kerr spacetime, the partial absorption cross section becomes negative for some values of $(\omega,m)$. 

Some of the features observed in the scalar absorption by Kerr BH have also been observed in the absorption of sound waves by the draining bathtub: an (inexact) analogue of Kerr BH in (2+1)-dimensions \cite{Oliveira:2010zzb} which is amenable to a full analysis using the complex angular momentum approach \cite{Dolan:2011ti}. The results presented here represent a significant step towards understanding the absorption by axially-symmetric BHs, for waves impinging at general angles. Possible themes for future work could include (i) an extension of the complex angular momentum approach of Sec.~\ref{subsec:sinc} to waves impinging at arbitrary angles of incidence, which would require careful asymptotic analysis of the spheroidal harmonics, and (ii) analysis of higher-spin (e.g.~Dirac \cite{dolanthesis}, electromagnetic, or gravitational) planar waves, where there will be an additional coupling between black hole rotation and the spin of the field.

\begin{acknowledgments}
We would like to acknowledge Conselho Nacional de Desenvolvimento Cient\'ifico e Tecnol\'ogico (CNPq), Coordena\c{c}\~ao de Aperfei\c{c}oamento de Pessoal de N\'ivel Superior (CAPES) and Marie Curie action NRHEP-295189- FP7-PEOPLE-2011-IRSES for partial financial support. LC acknowledges support from the Abdus Salam International Centre for Theoretical Physics through the Associates Scheme. SD is grateful to Huan Yang for discussions, exchanges of notes, and validation of numerical results.
\end{acknowledgments}

\bibliography{refs_abs}
\end{document}